\documentclass{egpubl}
\usepackage{eg2021}
 
\ConferencePaper        
\usepackage[T1]{fontenc}
\usepackage{dfadobe}  

\biberVersion
\BibtexOrBiblatex
\usepackage[backend=biber,bibstyle=EG,citestyle=alphabetic,backref=true]{biblatex} 
\electronicVersion
\PrintedOrElectronic

\ifpdf \usepackage[pdftex]{graphicx} \pdfcompresslevel=9
\else \usepackage[dvips]{graphicx} \fi

\usepackage{egweblnk} 

\usepackage{color}

\definecolor{blue}{rgb}{0,0,0.6}
\definecolor{green}{rgb}{0,0.3,0}
\definecolor{red}{rgb}{0.6,0,0}
\definecolor{gray}{rgb}{0.4,0.4,0.4}
\definecolor{black}{rgb}{0,0,0}
\definecolor{lightgray}{rgb}{0.83, 0.83, 0.83}
\definecolor{purple}{rgb}{1,0,1}

\usepackage{graphicx}
\usepackage{multirow}
\usepackage{amsmath}
\usepackage{gensymb}
\usepackage[ruled,vlined]{algorithm2e}
\usepackage{graphicx}
\usepackage{float}
\usepackage{makecell}
\usepackage{scalerel,stackengine}
\usepackage{cancel}
\stackMath
\newcommand\reallywidehat[1]{%
\savestack{\tmpbox}{\stretchto{%
  \scaleto{%
    \scalerel*[\widthof{\ensuremath{#1}}]{\kern.1pt\mathchar"0362\kern.1pt}%
    {\rule{0ex}{\textheight}}
  }{\textheight}%
}{2.4ex}}%
\stackon[-6.9pt]{#1}{\tmpbox}%
}
\newcommand\inv[1]{#1\raisebox{1.15ex}{$\scriptscriptstyle-\!1$}}
\newcommand{\norm}[1]{\left\lVert#1\right\rVert}

\begin{document}

\title[LoBSTr: Real-time Lower-body Pose Prediction from Sparse Upper-body Tracking Signals]%
{LoBSTr: Real-time Lower-body Pose Prediction from Sparse Upper-body Tracking Signals}

\author[Yang et al.]
{\parbox
{\textwidth}
{\centering Dongseok Yang\orcid{0000-0002-4696-3465},
        Doyeon Kim\orcid{0000-0002-4442-314X}, and Sung-Hee Lee\orcid{0000-0001-6604-4709}}
\\
{\parbox
{\textwidth}
{\centering 
Korea Advanced Institute of Science and Technology (KAIST)\\}}}

%

\teaser{
\centering
\includegraphics[width=\linewidth]{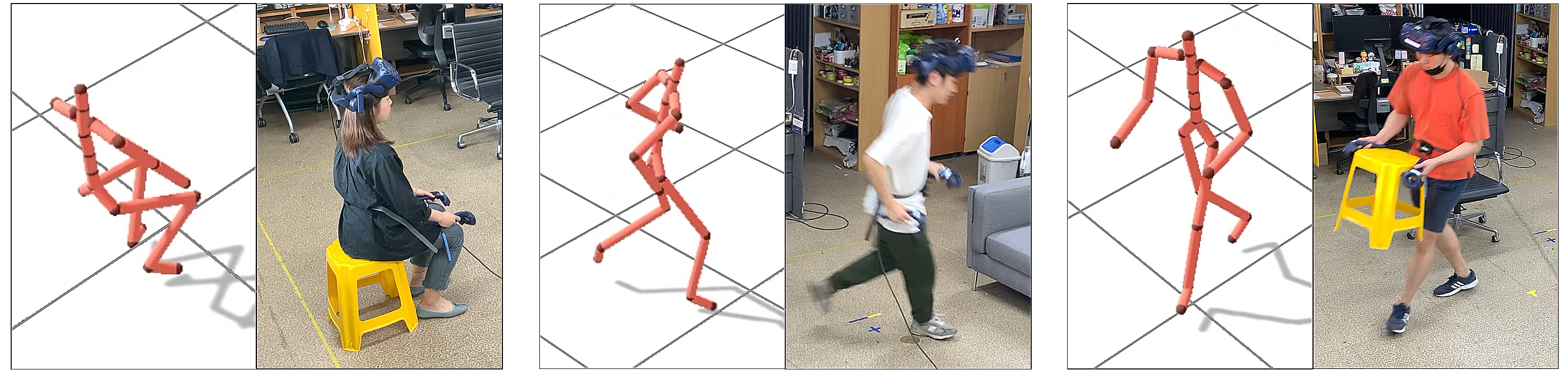}
\caption{Our system predicts the lower-body pose of the user from sparse tracking signals of the head, hands, and pelvis for a wide range of actions using off-the-shelf VR devices.}
\label{fig:teaser}
}

\maketitle
\begin{abstract}

With the popularization of games and VR/AR devices, there is a growing need for capturing human motion with a sparse set of tracking data. In this paper, we introduce a deep neural network (DNN) based method for real-time prediction of the lower-body pose only from the tracking signals of the upper-body joints. Specifically, our Gated Recurrent Unit (GRU)-based recurrent architecture predicts the lower-body pose and feet contact states from a past sequence of tracking signals of the head, hands, and pelvis. A major feature of our method is that the input signal is represented by the velocity of tracking signals. We show that the velocity representation better models the correlation between the upper-body and lower-body motions and increases the robustness against the diverse scales and proportions of the user body than position-orientation representations. In addition, to remove foot-skating and floating artifacts, our network predicts feet contact state, which is used to post-process the lower-body pose with inverse kinematics to preserve the contact. Our network is lightweight so as to run in real-time applications. We show the effectiveness of our method through several quantitative evaluations against other architectures and input representations with respect to wild tracking data obtained from commercial VR devices.


\end{abstract} 
\section{Introduction}

Human motion capture is gradually expanding its application areas. 
While precise motion capture with a dense set of sensory data is widely adopted for movie making, there is a growing need for capturing human motion with only a sparse set of data for mass-market, such as game and VR/AR, due to cost and usability reasons. 

A reasonably minimal number of tracking points for the whole body motion capture would be six: head, hands, feet, and trunk. The whole body motion can then be reconstructed based on inverse kinematics.
However, many commercial body-worn trackers (e.g., HTC VIVE and Oculus Rift) have limitations, especially for feet tracking.

First, trackers are vulnerable to impacts between the feet and floor during locomotion. The impact causes instantaneous deviation of the trackers from the body, which induces a large sensory error. In addition, feet trackers have a high probability of failure due to the occlusion by nearby objects if the capture space is equipped with furniture and objects.

Except for the motions in which leg movement is important (e.g., kicking), the leg generally takes the role of moving the upper-body to some desired direction or supporting the posture of the upper-body. For this range of motions, is it possible to estimate the natural motion of the lower-body with only the sparse tracking data of the upper-body? Will such a lower-body motion estimator be valid in a wide range of motions, from locomotion that shows high correlations between upper and lower bodies to the motions in which such correlation is weak (e.g., in-place upper-body movement)?

To address these questions, we develop a novel deep neural network (DNN)-based architecture, named \emph{LoBSTr} (\textbf{Lo}wer-\textbf{B}ody prediction with \textbf{S}parse \textbf{Tr}ackers), that predicts lower-body pose given the tracked position and orientation data for the head, hands, and pelvis. The goal of our network is challenging as it is severely underdetermined; sparse tracking information of upper-body joints can be mapped to many different lower-body poses, which are also affected by the body size and proportion. 
To tackle this challenge, we carefully designed the input representation and network structure, which are the major contributions of our work. First, to reduce the ambiguity of mapping from upper-body signals to lower-body pose, LoBSTr takes as input past temporal sequence of upper-body signals. Its recurrent neural network (RNN)-based architecture then outputs the lower-body pose at the current frame. Second, to model the correlation between upper-body and lower-body motions and increase the robustness against the diverse sizes and proportions of the user body, the input signal is represented in terms of the velocity. In addition, to remove foot-skating and floating artifacts, LoBSTr predicts feet contact states, which are used to post-process the lower-body pose with inverse kinematics to preserve the contact. Lastly, our network is designed to be lightweight so as to run in real-time applications (45 fps, < 22ms).

As a result, our method can generate plausible lower-body motions in a reasonable range for VR/AR applications from locomotion to those which have weak correlations with the upper body, e.g., upper-body gesture while walking or standing, allowing for a significantly wider range of capturable motions than previous 3-point tracking systems \cite{deepmotion3point, lin2019temporalik} (Figure \ref{fig:teaser}). 

By removing feet trackers for full-body avatar motion generation, our system is free from artifacts caused by foot-floor impact and tracker occlusion in a cluttered environment, achieves a larger tracking area than those of conventional systems, and reduces the cost for devices; thereby making full-body tracking more accessible to the general VR/AR users.

In addition, the robustness of our method allows us to train the model with a single body size, using existing motion capture datasets such as CMU \cite{cmu2013}, PFNN \cite{holden2017phasefunctioned}, and MHAD \cite{ofli2013berkeley} datasets, yet to be applicable to different body proportions tracked with commercial trackers. Our network architecture successfully learns temporal features of human motions to reconstruct different motions and transitions between them without phase or contact labeling.

Our method is evaluated quantitatively by comparing average positional and rotational errors of different network architectures. Different input representations are evaluated in terms of toe-base distance error and average body movement \cite{starke2020localphase} to assess the model's robustness to user body shape change. An ablation study over loss terms is performed to see the effects of proposed training loss terms. A qualitative comparison is also conducted against a baseline 6-point tracking system.
\section{Related Work}

This section introduces previous work related to ours with respect to the problems of obtaining reduced space for human motion and estimating human motion from sparse sensor signals.

\subsection{Motion Synthesis from Reduced Space Representations}

Obtaining a reduced space representation of human motion that captures only the valid scope of human motions enables generating natural motions as well as editing and optimizing motions faster. 

Early methods used linear dimensionality reduction models. Chai and Hodgins \cite{chai2005performancelowdim} applied the principal component analysis (PCA) to construct locally linear pose spaces online and synthesized natural human pose with low dimensional control inputs. Liu et al.  \cite{liu2006humanreducedmarker} proposed a PCA-based clustering method to map the full set of marker configurations to a lower-dimensional set while preserving the original captured poses. Liu et al. \cite{liu2011realtimeinertial} constructed a dynamic linear motion space online by searching a set of similar motion clips from the dataset and estimating the current pose under the maximum a posteriori (MAP) framework. Safonova et al. \cite{safonova2004synthesizinglowdim} synthesized realistic human motions by applying physics constraints to adjust poses obtained by PCA-based methods. Linear methods are simple and fast but can only model a narrow range of behavior-specific motions. 

Nonlinear reduction methods can better represent human motion by reflecting the nonlinear characteristics of human motion.
Grochow et al. \cite{grochow2004styleik} and Levine et al. \cite{levine2012continuouslowdimcontrol} used Gaussian Process Latent Variable Model (GPLVM) to obtain a low dimensional latent space to reconstruct high dimensional motion data. The nonlinear models are generally effective when dealing with a small number of motion samples but their space and time complexities grow fast with the size of learned motion data.

Recently, DNN-based methods are developed to obtain a latent space for synthesizing and reconstructing motions. Holden et al.  \cite{holden2016deeplearningframework} used a convolutional autoencoder to obtain a motion manifold, which is used to synthesize and edit motions with high-level control inputs such as locomotion paths. Jang and Lee \cite{jang2020constructingsequentialmanifold} developed an RNN-based method to learn a motion manifold trained with a sequence-to-sequence architecture. Ling et al. \cite{ling2020character} developed autoregressive conditional variational autoencoders that learn reduced space of plausible human motions for character motion generation. \cite{holden2020learned} applied DNN to the motion matching technique to reduce memory usage while retaining the quality of the controllable motion.
Our work also learns a reduced motion space using a DNN architecture. However, unlike previous work that reduces the whole-body data to generate whole-body motions, we obtain a reduced space from the upper-body data to generate the corresponding lower-body motions.

\subsection{Real-time Pose Prediction from Sparse Tracking Signals}

The task of predicting full-body pose from sparse signals is an underdetermined problem, where many different poses can satisfy the given inputs. To obtain the most plausible pose among them, researchers have developed methods that leverage additional knowledge or heuristics on human motion or learn to predict plausible poses from examples.

Recently deep learning approaches are used for pose prediction and show promising results. Wouda et al. \cite{wouda2019timefullbodyfiveiner} proposed a bi-directional Long Short-Term Memory (LSTM) model to estimate full-body pose from 5 IMUs and showed that their method is superior to a baseline shallow learning method with respect to preserving temporal coherency of motion. Huang et al. \cite{huang2018deepinertialposer} developed a bi-directional RNN to learn the temporal pose priors and reconstructed human pose from 6 Inertial Measurement Units (IMUs) worn on the body. They first trained the network with synthetic IMU data and then fine-tuned it with a real dataset.
These approaches predict full-body pose given sparse signals from sensors attached to all limbs and other body parts, whereas our method aims to predict the lower-body from upper-body tracking signals.

Retrieving full-body pose from sparse observations is also actively researched in computer vision. Cheng et al. \cite{cheng2019occlusion} used 2D and 3D temporal convolutional networks (TCNs) to reconstruct 3D human pose from occluded monocular video. \cite{rockwell2020full} showed that  3D mesh reconstruction can be improved by pre-training networks with cropped images in an annotated dataset.

Several studies proposed methods to predict the lower-body pose from the upper-body information. Jiang et al. \cite{jiang2016realfullbodyofftheshelf} developed a method to generate lower-body motion by blending 8 walking animations with different directions based on the body direction and velocity, which are predicted from off-the-shelf VR devices. As a blended motion, the resulting motion looks natural but its scope is limited by the predefined animations. A balance control-based method \cite{thomasset2019lowerlowerbodycontrolzmp} reconstructs static pose and locomotion of the lower-body according to the tracked upper-body joints. Specifically, the target Zero Moment Point (ZMP) trajectory is determined from the upper-body motion, and the full-body animation is generated to realize the ZMP trajectory. \cite{deepmotion3point} introduced a commercial 3-point tracking system with a physics-based character model, trained to satisfy physical and kinematic constraints using deep reinforcement learning. The resulting motion may take unnatural steps, different from the actual motion of the user. Recently, Lin \cite{lin2019temporalik} developed an LSTM-based pose estimator with 3 trackers worn on the user's head and hands. The trained model successfully estimates the upper-body pose but often fails to predict the lower-body pose.
Compared with these studies, our method is capable of generating more natural lower-body motions for a wider range of actions.
\section{Prediction of the Lower-Body Pose}

\begin{figure*}[t]
  \centering
  \includegraphics[width=\linewidth]{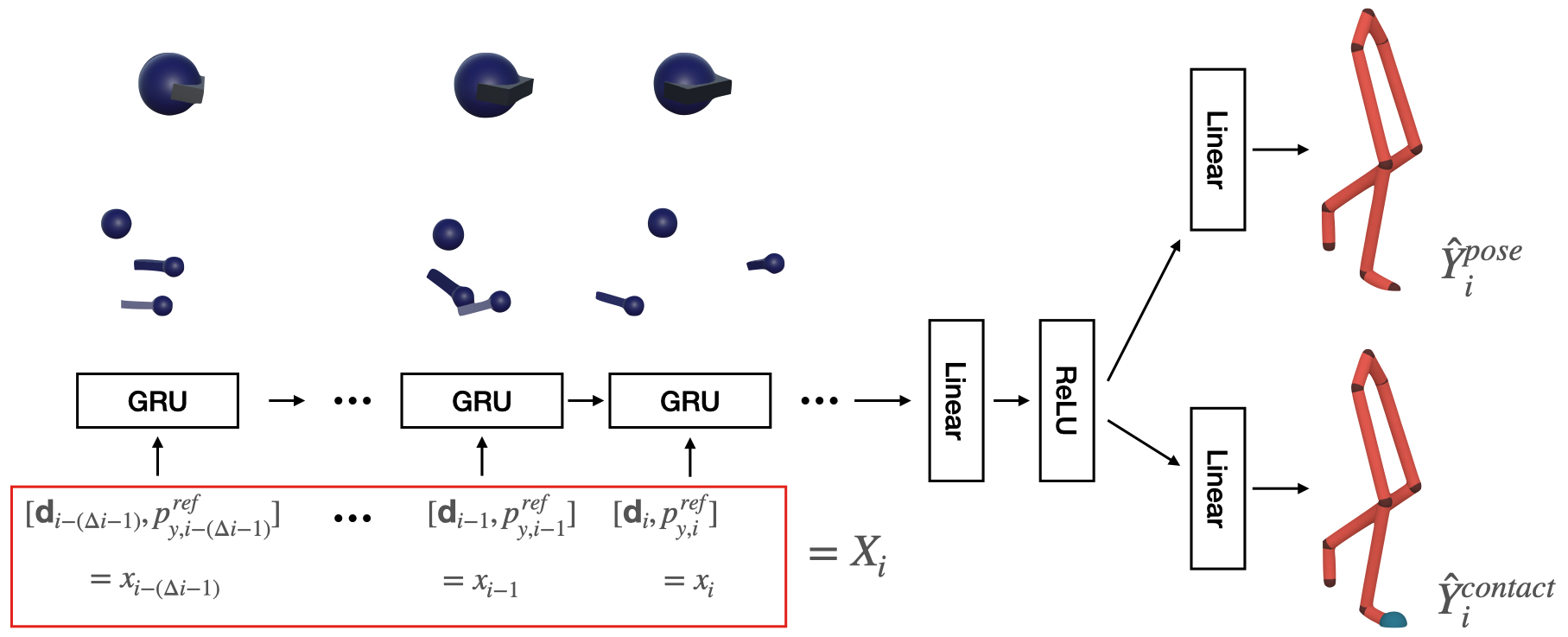}
  \caption{\label{fig:Netarch1}Our LoBSTr network architecture to predict the lower-body pose given a sequence of upper-body tracking signals.}
\end{figure*}

Figure \ref{fig:Netarch1} shows the overview of our LoBSTr network architecture.
It infers the lower-body pose and feet contact states at the current frame from a past sequence of tracking signals for the 4 upper-body joints, including the head, hands, and pelvis.
The input sequence is mapped to a latent representation by an encoder and passed to two linear layers, one for lower-body pose prediction and the other for feet contact prediction. The output pose is post-processed with respect to the output contact state by inverse kinematics for better visual quality.

\paragraph*{Pose Representation.}
We first describe the representation of pose in our work.
A pose of a character $\textbf{q}_i$ at frame $i$ is represented with the world position $p^{root}$ and orientation $q^{root}$ of the root $\textbf{r}_i = [p_{i}^{root}, q_{i}^{root}]$ and the local rotations of remaining joints $\textbf{j}_i = [q_{i}^{k}]_{k=1}^{n_{joint}}$, i.e., $\textbf{q}_i = [\textbf{r}_i, \textbf{j}_i]$. All rotations and orientations are represented in 6-DoF with forward (Z-axis) and up (Y-axis) vectors.
To express motion in the user's egocentric space, we define a virtual reference joint, of which frame is located at the root joint and oriented to point to the world up vector with its Y-axis and to the frontal direction with its Z-axis. The frontal direction is obtained by projecting the frontal direction of the pelvis joint to the ground plane.

\subsection{Velocity-based Prediction}

Predicting pose from the position and orientation data of trackers is highly sensitive to the user's body shape, which causes variation in tracker configurations. For example, the pelvis tracker tends to be attached farther from the actual pelvis joint for an overweight user and has different initial rotations for every run; it requires careful calibration on tracker position and orientation to generate natural poses from those raw signals. To avoid this, we designed our input representation to be least affected by the variation of tracker configurations and user's body shape but to be directly acquired from the raw tracking signals.

To this end, we represent the input motion of the 4 joints (i.e., the head, hands, and reference joints) in terms of the velocities. Specifically, the velocities of sparse joints are described with respect to the defined reference coordinate frame, which is different from conventional body velocities of full-body joints \cite{chiu2019action}. We argue that this velocity representation is more robust to differences in tracker configuration and body shape of the users than position and orientation representation, and thus allows our network to produce robust output from different users with a simple calibration step.

The velocities $\textbf{d}_{i}$ of tracked joints at the $i^{th}$ frame consist of the linear and angular velocities $d_{i}^{ref}=[v_{i}^{ref}, w_{i}^{ref}]$ of the reference joint and the linear and angular velocities $d_{i}^{joint}=[v_{i}^{joint}, w_{i}^{joint}]\in R^{(3+6)}$ of the three joints, expressed with respect to the current reference joint frame. Refer to Appendix 1 for the equations to compute $\textbf{d}_{i}$.

The input $X_i$ to our network at the $i^{th}$ frame includes the velocity vector $\textbf{d}$ and the height $p^{ref}_{y}$ of the reference joint above the ground, in the frame range of $[i-(\Delta{i}-1), i]$:
\begin{align}
X_i  & = [x_{i - ({\Delta}i-1)},\ x_{i - ({\Delta}i-2)},\ \ldots\ ,\ x_{i}]\in R^{\Delta{i} \times \dim(x)} \\
x_i  & = [\textbf{d}_{i}, p^{ref}_{y,i}]\in R^{(4 \times 9 + 1)} \\
\textbf{d}_{i} &= [d_{i}^{ref}, [d_{i}^{joint, k}]_{k=1}^{3}]
\end{align}

Then our network outputs the information to predict the lower-body pose at the current frame $i$.
These are the local rotations of 8 lower-body joints (the hip joints, upper-legs, lower-legs, and feet) ${\hat{Y}_{i}}^{pose} \in R^{(8 \times 6)}$ and contact probabilities of feet ${\hat{Y}_{i}}^{contact} \in R^{(2 \times 2)}$. Feet contact states are determined from the output contact probabilities and used to remove the foot-skating artifact in the post-processing step.

\subsection{Network Architecture}

We use an RNN with Gated Recurrent Units (GRUs) \cite{chung2014empiricalgru} as the latent encoder model, which has shown good performance in preserving temporal continuity of human motions \cite{wouda2019timefullbodyfiveiner}. GRUs have the remember-forget function that is capable of learning implicit priorities over frames to form a hidden space from the input frame sequence. Furthermore, GRU structure requires a fewer number of parameters than LSTM \cite{hochreiter1997longshorttermmem} to achieve similar performance, thereby allowing for faster computation time, an important quality for real-time applications.

In our network architecture, one layer GRU network with a hidden dimension of $1024$ encodes the input vectors to a latent representation, conditioned by the input vectors of previous frames. Specifically, an input vector $x_{i}$ of the $i^{th}$ frame is encoded as follows: $h_{i}= GRU(h_{i-1},x_{i}|\theta_g)$, where $h_{i-1}$ is the hidden state of the $({i-1})^{th}$ frame and $\theta_{g}$ is the hidden layer parameter. After the input vector $x_{i}$ at the current frame is encoded to $h_{i}$, $h_{i}$ passes through a linear layer with parameter $\theta_{l_1}$ and ReLU activation, generating a latent vector of $128$ dimensions.
Finally, the latent vector passes through two linear layers with parameters
$\theta_{l_p}$ and $\theta_{l_c}$ to produce the local rotations of the 8 lower-body joints ${\hat{Y}_{i}}^{pose}$ and the contact probability ${\hat{Y}_{i}}^{contact}$,  respectively.
In our experiment, the dimensions of each parameter are as follows: $\dim(\theta_{g})=(4\times9 +1)\times1024$, $\dim(\theta_{l_1})=1024\times128$, $\dim(\theta_{l_p})=128\times(8 \times 6)$, $\dim(\theta_{l_c})=128\times(2 \times 2)$.

The size of the time window $\Delta{i}$ is set to 45 (= 1 second for 45 fps data), which was found empirically to be appropriate to capture characteristics of different motion categories and maintain temporal continuity of output lower-body poses when played sequentially in real-time.

\subsection{Network Training}

We train the network in a supervised manner.
In order to include various actions ranging from locomotion to in-place upper body motion, we built a dataset combining parts of several motion capture datasets. Specifically, a training set of $(X_i,Y_i)$ is obtained from PFNN, MHAD, and CMU motion datasets and the network is trained to output as close as to $Y_i$ given $X_i$. Refer to Appendix 2 for the details on preparing and augmenting the motion dataset for the training.

Following the previous studies, we optimize our network to minimize four loss terms: pose loss, forward kinematics loss \cite{pavllo2018quaternet, villegas2018neural}, feet velocity loss \cite{pavllo2019modeling, kim2020motion}, and contact prediction loss \cite{shi2020motionet}. The pose loss is our target loss which directly affects the accuracy of the predicted lower-body pose. The remaining three terms work as regularizers to promote the naturalness of feet trajectories and reduce artifacts such as foot-skating and floating.

\textit{Pose loss.}
As predicting the lower-body pose at the current frame is the goal of our network, it is trained to minimize the $L^1$ norm between the predicted lower-body pose and the ground truth $Y_{i}^{pose}$.
\[\mathcal{L}_{pose} = \norm{{\hat{Y}_{i}}^{pose} - Y_{i}^{pose}}_1\]
        
\textit{Forward kinematics (FK) loss.}
Minimizing 3D positional errors of the feet results in better perceptual quality by decreasing skating and floating artifacts of the feet. The loss is defined as the $L^2$ norm between the toe-base joint positions computed by FK from the predicted and ground truth lower-body poses.
\[\mathcal{L}_{FK} = \norm{FK({\hat{Y}_{i}}^{pose}) -  FK({Y_{i}}^{pose})}_2\]

\textit{Feet velocity loss.} 
The jiggling of the resulting feet trajectory is one of the major artifacts which severely degrades the quality of output motion. We designed a velocity loss, which drives our network to produce the ground truth velocity of toe-base joints.
\[\begin{array}{ll}
\mathcal{L}_{velocity}& = \lVert(FK({\hat{Y}_{i}}^{pose}) -  FK(Y_{i-1}^{pose})) \\
& -(FK({Y_{i}}^{pose}) - FK(Y_{i-1}^{pose}))\rVert_2
\end{array}\]

\textit{Contact Prediction Loss.}
In addition to the lower-body pose, our model predicts the contact state of each toe-base joint from the latent vector of the input sequence. The contact prediction layer is a binary classification layer trained to minimize the binary cross-entropy between output label probability and ground truth contact label.  
\[\mathcal{L}_{contact} = CrossEntropy({\hat{Y}_{i}}^{contact}, Y_{i}^{contact})\]
        
Combining the four loss terms, the final loss function is
\[\mathcal{L}\ =\ \lambda_{1}\mathcal{L}_{pose}+\lambda_{2}\mathcal{L}_{FK}+\lambda_{3}\mathcal{L}_{velocity}+\frac{\lambda_{4}}{2}(\mathcal{L}_{contact}^{left} + \mathcal{L}_{contact}^{right}),\] where hyper-parameters for experiments are set as follows: $\lambda_{1}=1$, $\lambda_{2}=0.1$, $\lambda_{3}=0.1$, and $\lambda_{4}= 10^{-6}$.
Refer to Appendix 3 for the details on network training.

\begin{figure}[t]
  \centering
  \mbox{} \hfill
  \includegraphics[width=\columnwidth]{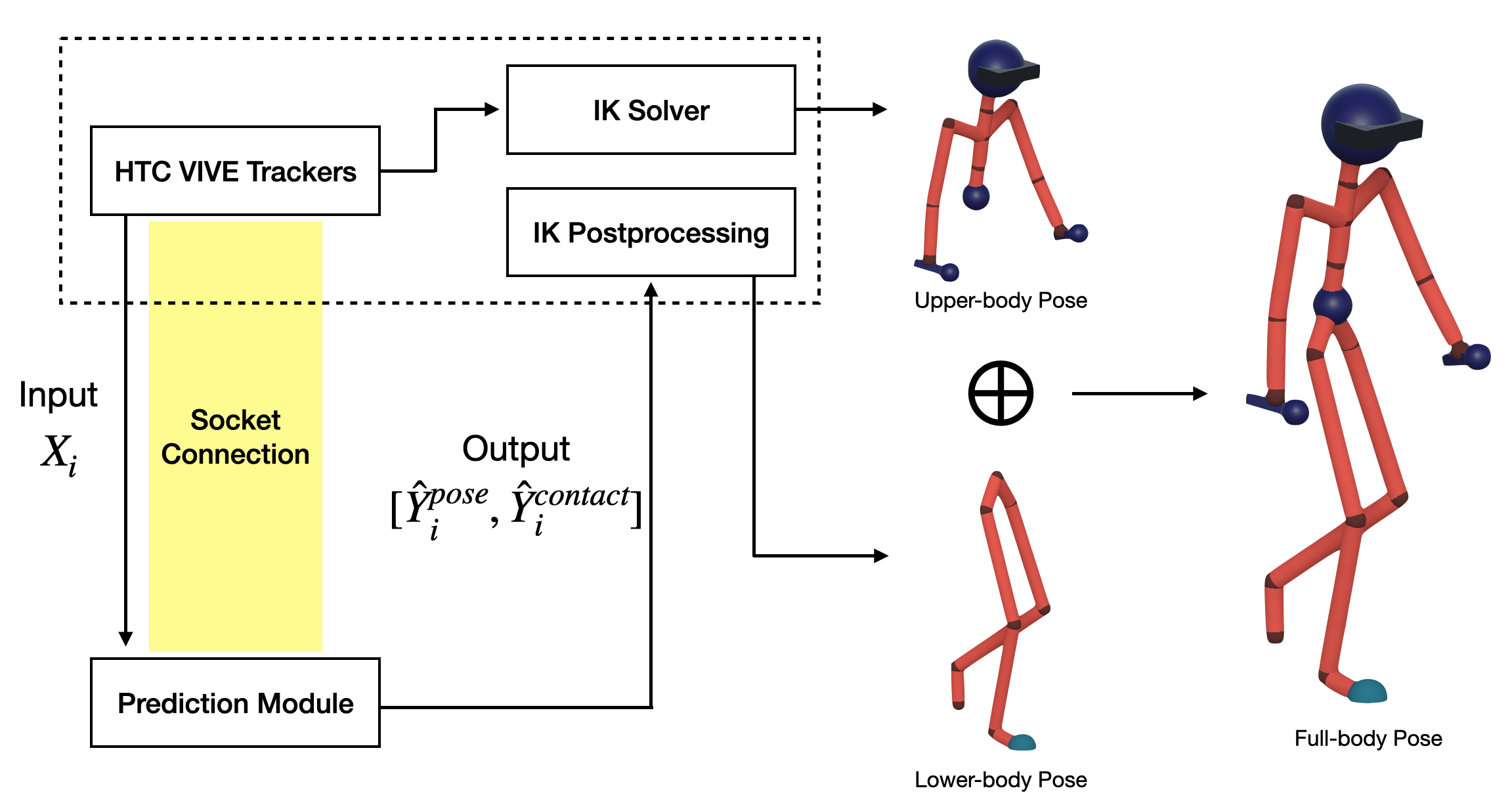}
  \hfill
  \mbox{}
  \caption{\label{fig:system}%
           Overview of our full-body avatar system.}
\end{figure}

\subsection{Real-Time Motion Generation}

By using our lower-body pose prediction network, we develop a real-time full-body avatar system that uses off-the-shelf VR trackers as shown in Fig. \ref{fig:system}.  Specifically, HTC Vive Pro set is used as a tracking device: HMD, two hand-held controllers, and a tracker on the pelvis. Input to the network is computed from the world transformations of trackers. Note that there may be a large deviation from the tracker positions and the joint positions. For example, the pelvis tracker worn in front of the waist is far from and the pelvis joint. However, we do not take any special treatment to compensate for this difference, which is challenging due to the user's shape variations. Instead, our input representation of velocities robustly handles the deviation.

Our system is implemented with Unity3D engine and SteamVR platform and runs at a fixed framerate of 45 fps. The system requires a warm-up time of 1 second to generate an initial input of 45-frames. The input sequence is sent to the lower-body pose prediction module every frame via TCP socket connection. The average inference time is around 2.5ms, which is fast enough to maintain the running rate of 45 fps. The upper-body pose is computed to match the tracked end-effector transformations by an IK solver \cite{root2017final} and combined with the lower-body pose to animate the virtual character skeleton.

\subsubsection{Contact Post-processing}

While our network outputs plausible poses of the legs, continuous animation of the leg shows some extent of foot-skating and floating artifacts. These artifacts are more visible when users take a vigorous motion like running or sharp turn, which induces high-frequency shaking of the pelvis tracker.

To resolve those artifacts that occur during runtime, we train our model to output contact probabilities of two toe-base joints and post-process the leg pose in the contact state. At the moment when the contact state changes from false to true, the toe-base position is set as the target contact position and a Jacobian-based inverse kinematics solver is used to maintain the contact while the contact state remains true. 

To maintain a smooth transition when the toe-base joint loses contact, a simple interpolation is performed for a fixed time window ($10$ frames in our experiment); the toe-base joint position at the contact-losing frame and that computed from the network output are linearly interpolated to determine the target toe-base joint position. The corresponding leg pose is computed from a Jacobian-based inverse kinematics solver. The interpolation parameters are computed from a shifted sine function to implement a slow-in-slow-out transition.
\section{Experiment}

We assess the effectiveness of our method through several offline and online evaluations as shown below.

\subsection{Offline Evaluation}

\begin{figure*}[t]
  \centering
  \mbox{}
  \includegraphics[width=\linewidth]{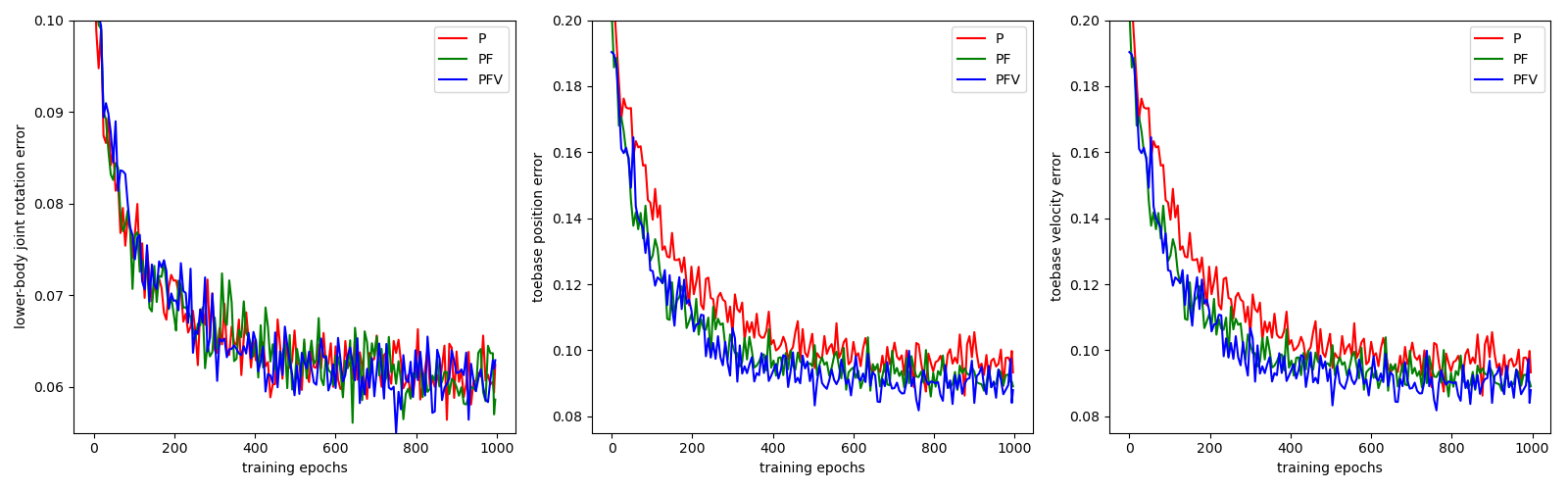}
  \caption{\label{fig:ablation} Errors of different loss term combinations according to training epochs.}
  \mbox{}
\end{figure*}

\subsubsection{Network Architecture Comparison}

To the best of our knowledge, our work is the first DNN-based method trained with existing motion capture datasets to predict lower-body pose only from the input sequence of upper-body joints. Therefore, we compare models with baseline network architectures of fully connected layers (FC), convolutional layers (CNN), and different types of recurrent unit layers (RNN, LSTM, and GRU).
We assess the ratio of correct contact labels out of the whole frames, average rotational error per joint (8 lower-body joints), and average positional error of two end-effectors (left and right toe-bases). The networks only vary in the latent mapping module for input sequence; the encoder is replaced with different network architectures while linear layers for predictions remain identical. Hidden and latent dimensions are fixed to 1024 and 128, respectively. The FC encoder consists of 6 layers of FC layers \cite{holden2018robust} and the CNN encoder consists of 1D convolution and max-pooling layers \cite{holden2015learning}.
Table \ref{tab:ablation} shows the errors of different architectures for the test dataset.

In addition, we tested a GRU encoder with an autoregressive structure, which was reported to improve motion prediction accuracy and reduce visual artifacts in the output sequence \cite{martinez2017human}.
Our tested autoregressive model forms a latent vector by concatenating a latent vector output from a GRU encoder, which takes the upper-body tracking signals as input, and a linear layer (parameter dimension of $48\times128$), whose input is the lower-body pose of the previous time step. This autoregressive model leads to the explosion of the resulting lower-body pose after approximately 120 frames of running, presumably because of the accumulation of errors in the lower-body pose.

The comparison shows that the models with recurrent structure and remember-forget function (LSTM and GRU) produce the best performance on all three measures and the difference between the two is not significant.
We choose to use GRU for its higher prediction accuracy for contacts as well as having fewer cell parameters, which allows for faster training and inference time.

\begin{table}[h]
\centering
\caption{Network architecture comparison on the test dataset.}
\begin{tabular}{|p{1.8cm}|p{1.5cm}|p{1.5cm}|p{1.5cm}|} 
\hline
Model & Contact Accuracy & Rotational Error & Positional Error \\
\hline
\hline
FC-6 & $84.22\%$ & $9.37\degree$ & $7.2$cm \\
CNN-1D & $72.33\%$ & $13.86\degree$ & $13.3$cm \\
RNN & $82.27\%$ & $9.43\degree$ & $7.1$cm \\
LSTM & $83.85\%$ & $8.50\degree$ & $6.63$cm \\
GRU (Ours) & $\textbf{85.77}\%$ & $8.53\degree$ & $6.63$cm \\
\hline
\end{tabular}
\label{tab:ablation}
\end{table}%

In addition to the quantitative analysis, we evaluate the visual quality of the output animation. Architectures with recurrent units (RNN, LSTM, and GRU) produce smooth and continuous motions while others suffer from jittery output motions.

\subsubsection{Loss Term Ablation Study}
We examine the effect of proposed loss terms ($\mathcal{L}_{FK}$ and $\mathcal{L}_{velocity}$) by doing an ablation study for the terms. The proposed model of GRUs is trained by three different loss term combinations: 1. $\mathcal{L}_{pose}$, 2. $\mathcal{L}_{pose} + \mathcal{L}_{FK}$, 3. $\mathcal{L}_{pose} + \mathcal{L}_{FK} + \mathcal{L}_{velocity}$, denoted as \textbf{P}, \textbf{PK}, and \textbf{PKV}, respectively. Figure \ref{fig:ablation} shows that adding $\mathcal{L}_{FK}$ and $\mathcal{L}_{velocity}$ does not affect the joint angle error but decreases the average position and velocity errors of the feet. Table \ref{tab:positionErrors} shows that the average positional error of the feet is minimum with \textbf{PKV}. We evaluate the visual quality of output lower-body motion from the models trained with the three loss term combinations, among which \textbf{PKV} gives significantly more stable lower-body motions, especially when the user is performing in-place upper-body motions.

\begin{table}[h]
\centering
\caption{Errors of different loss term combinations on the test dataset.}
\begin{tabular}{|l||c|c|c|} 
\hline
Loss Term Combination & P & PK & PKV   \\
\hline
Positional Error (cm) & $6.95$  & $6.64$  & $\textbf{6.63}$ \\ 
\hline
\end{tabular}
\label{tab:positionErrors}
\end{table}%

\subsubsection{Input Representations Comparison}

The robustness of the proposed velocity input feature is evaluated by two measures. First, we measure toe-base distance error, the difference of the distances between two toe-bases from the output pose and ground truth pose captured with trackers. Secondly, we compare body movement \cite{starke2020localphase}, the sum of absolute joint angle updates in output pose, generated from recorded inputs. Body movement metric examines whether the network is overfitted and produces over-smoothed motion for sub-domains represented by sparse training samples. The test set of wild signals is captured from subjects with 157, 171, and 184 cm heights. The subjects are asked to perform a sequence of actions; walk, run, sit/stand up, carry and move, static gestures, and free movement. The free movement contains motion categories that are not in training sets (e.g., walking backward, cross-steps, and jumping). Three identical models are trained with inputs of position-orientations, velocities, and a combination of both representations. For position-orientation representation, the input for reference joint is given in velocity representation.

\begin{table*}[t]
\centering
\caption{Average toe-base distance errors (cm) for users with different heights.}
\begin{tabular}{|c|c|c|c|c|c|c|c||c|} 
\hline
User Height & Representation & Walk & Run & Sit/Stand up & Carry and move & Static gestures & Free & \textbf{Total} \\
\hline
\hline
\multirow{3}{*}{157 cm} & velocity & 4.8 & 1.6 & 3.5 & 5.3& 4.0 & 9.5 & \textbf{4.4}\\
& pos-ori & 6.9 & 4.2 & 1.7 & 7.5& 0.2 & 11.0& 6.0 \\
& combined & 17.0 & 10.7 & 6.3 & 13.1 & 5.2 & 20.5 & 13.3 \\
\hline
\multirow{3}{*}{171 cm} & velocity & 3.7 & 1.3 & 1.3& 8.3& 0.7 & 7.5& \textbf{4.6}\\ 
& pos-ori & 8.3 & 5.3 & 2.1 & 6.3& 2.3 & 7.4& 5.0\\
& combined & 20.8 & 12.2 & 6.2 & 10.6 & 1.2 & 17.0 & 11.9 \\
\hline
\multirow{3}{*}{184 cm} & velocity & 0.03 & 1.4 & 2.4& 7.0& 9.2 & 3.3& \textbf{2.1}\\ 
& pos-ori & 2.9 & 1.2 & 1.6 & 7.7& 14.4 & 6.4& 5.2\\ 
& combined & 14.6 & 12.9 & 9.7 & 17.8 & 2.3 & 16.3 & 13.0 \\
\hline
\end{tabular}
\label{tab:toebaseDistance}
\end{table*}%

Table \ref{tab:toebaseDistance} shows the average toe-base distance error per frame over different user heights. For the whole sequence, velocity representation outperforms the other representations in average error for every subject.
When divided into motion categories, position-orientation representation shows smaller errors for a majority of the subjects (2 out of 3), only for the sit/stand up category.
The combined representation shows significantly larger errors for VR tracker data, although it outperformed the others for motion capture data. This result suggests that the model with combined input is trained to fit the distribution of the training domain (motion capture data) more tightly and thus becomes more sensitive to the body dimensions, tracker configurations, and wild noise than the velocity or position-orientation representation. From the result, we conclude that velocity representation generates much closer leg swings to the ground-truth than position-orientation representation.

\begin{figure}[t]
  \centering
  \includegraphics[width=\linewidth]{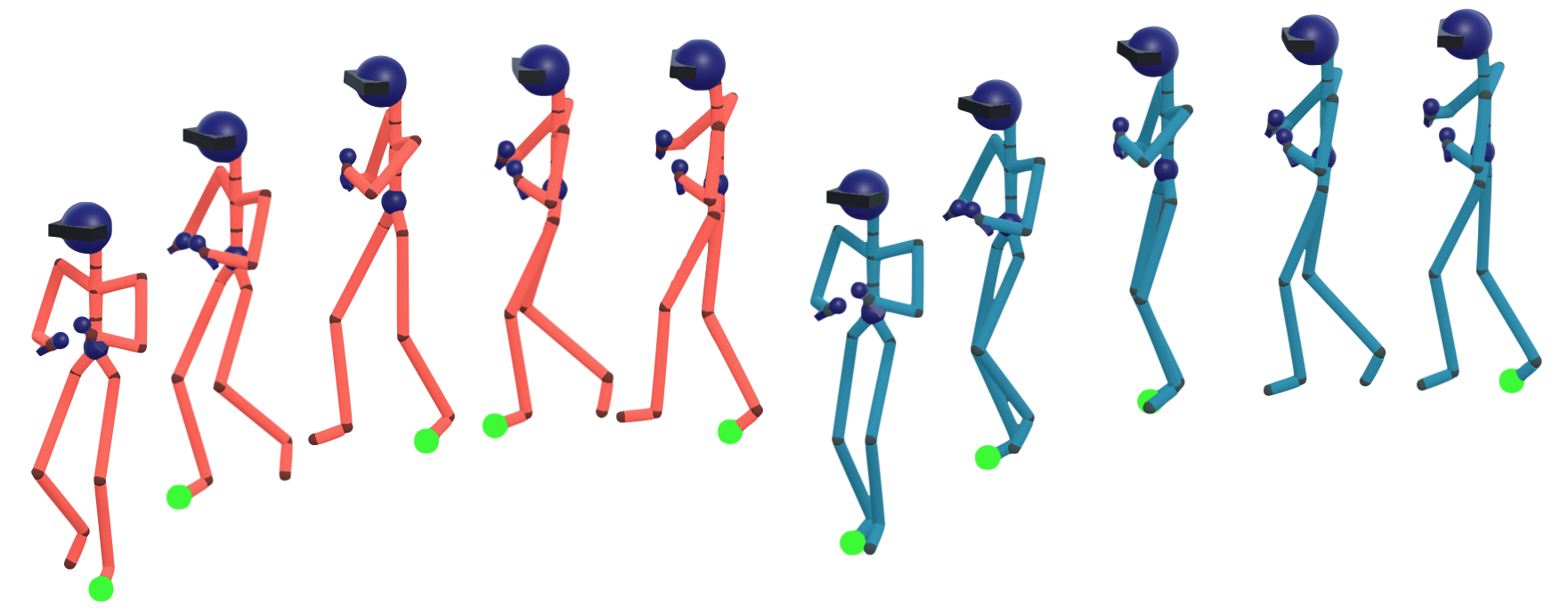}
  \caption{Styled running (arms swinging in front of the chest) motion by a 157cm subject. Compared with velocity input (red), position-orientation input (green) occasionally generates static poses. }
  \label{fig:157cm}
\end{figure}

\begin{figure}[t]
  \centering
  \includegraphics[width=\linewidth]{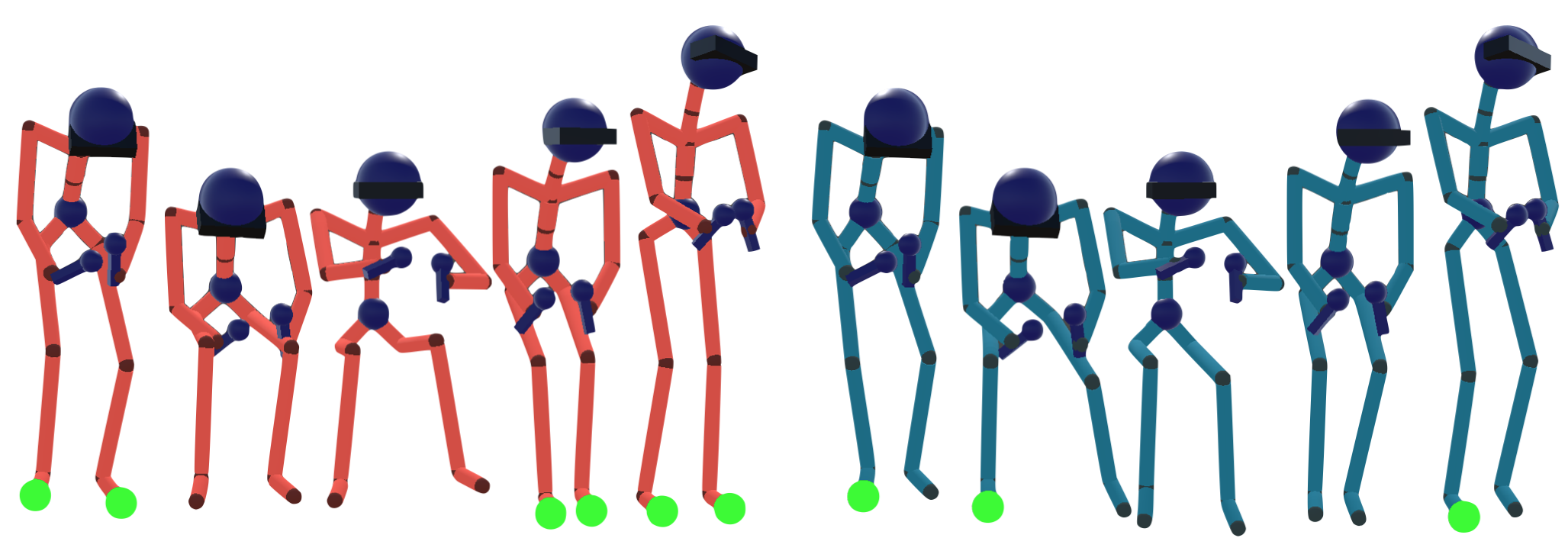}
  \caption{Sitting-standing up motion by a 176cm subject. Compared with velocity input (red), position-orientation input (green) shows less accurate contact state prediction.}
  \label{fig:176cm}
\end{figure}

\begin{figure}[t]
  \centering
  \includegraphics[width=0.8\linewidth]{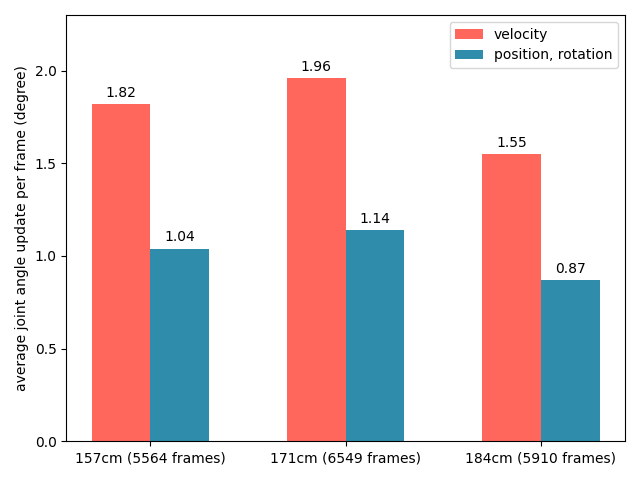}
  \caption{Average per-frame joint angle update of output poses from  users with different heights.}
  \label{fig:jointAngleUpdate}
\end{figure}

Figure \ref{fig:jointAngleUpdate} shows that velocity representation surpasses position-orientation representation in body movement measure for all users. While comparing visual quality of output lower-body motions, we found that the model with velocity representation learned larger motion space including sharp motions but that with position-orientation representation was overfitted to a smaller range of locomotion and sitting motion.

We observe that velocity representation generates continuous lower-body motion for the recorded test inputs, while position-orientation representation sometimes produces unnatural static poses during motion (Figure \ref{fig:157cm}). In addition, the model maintains higher precision of contact state prediction with the velocity representation (Figure \ref{fig:176cm}). From the results, we conclude that velocity representation outperforms position-orientation representation with respect to the robustness to unobserved input data (e.g., different body size and motion style), the accuracy of feet contact prediction, and visual quality.

\subsection{Online Evaluation}

We tested our real-time full-body avatar system for a wide range of motions performed by several participants with varying heights. Recordings of avatar and participant motions are provided in the supplemental video.

Before capturing, each participant takes T-pose in the calibration stage to compute rotational offsets between trackers and corresponding joints in the skeleton. Rotational offsets are multiplied to tracking signals to compensate for individual differences in tracker configuration, which may result from different controller grip and different wearing positions of trackers; the offsets not only vary among users but also for the same user at each session. Input joint velocities are then computed from the tracked world positions and orientations, compensated by the offsets. In addition, a height scale is obtained as the ratio of the pelvis heights between the participant and the standard skeleton used for the network training. The height of the participant's reference is multiplied by the height scale to map to the standard skeleton before feeding to the prediction network. Then our system outputs the whole-body pose of the standard skeleton, which is visualized after scaling with the height scale to reflect the participant's height.
 
A total of four people with heights of 157, 171, 180, and 184 cm participated in the test. The participants are first asked to perform basic actions, including walking, running, sitting, standing up, and static gestures. After that, they are asked to do actions that are not in the training set, including walking backward, moving objects to another place, and walking with greetings. Finally, they are asked to move freely with any actions they want to test. The capture environment is designed to represent the play area of general users; the room is 3.6m x 3.6m and is equipped with objects and furniture.

\begin{figure*}[p]
  \centering
  \includegraphics[width=\linewidth]{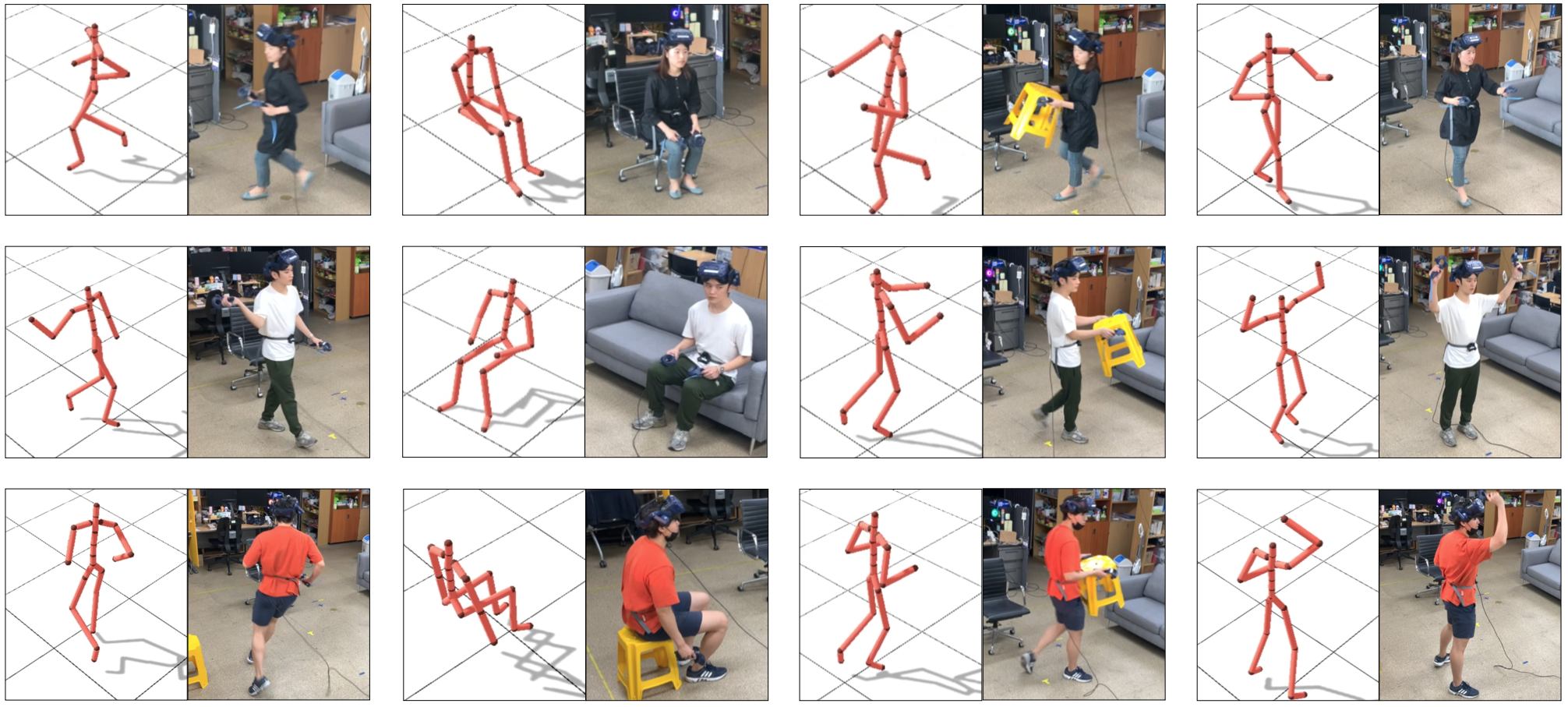}
  \vspace*{-3mm}
  \caption{\label{fig:DiffPeople}
           Avatar motions generated in real-time from users with various heights, 157cm (top), 171cm (middle), and 180cm (bottom).}
  \vspace*{13mm}

  \includegraphics[width=\linewidth]{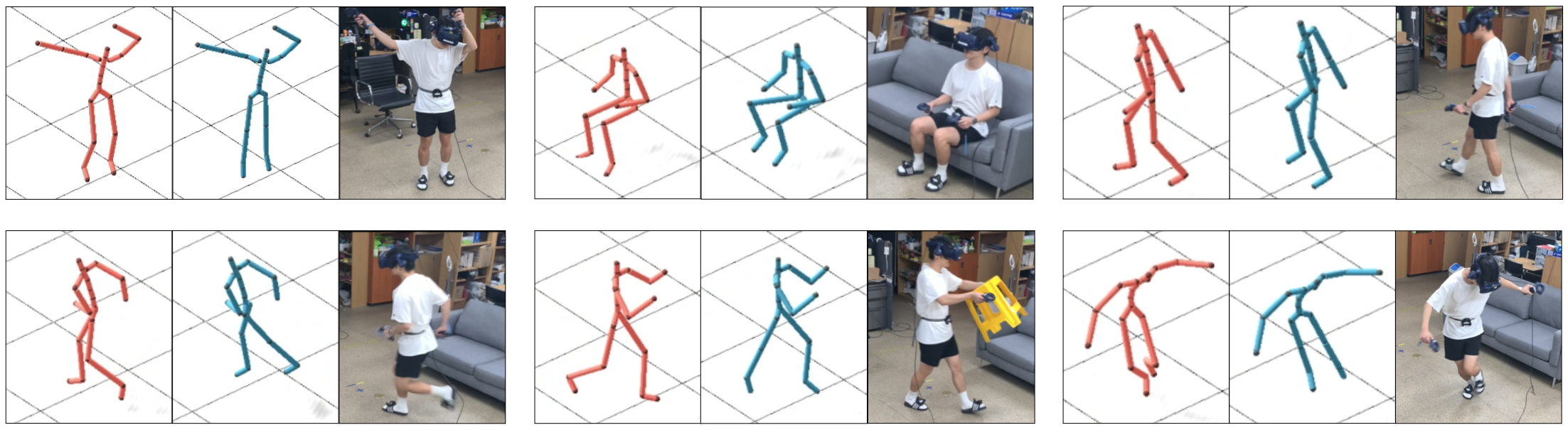}
  \vspace*{-3mm}
  \caption{\label{fig:Realtime}
           Comparison of real-time avatar motion against 6-point tracking (Red: Ours, Green: 6-point tracking).}
  \vspace*{13mm}

  \includegraphics[width=\linewidth]{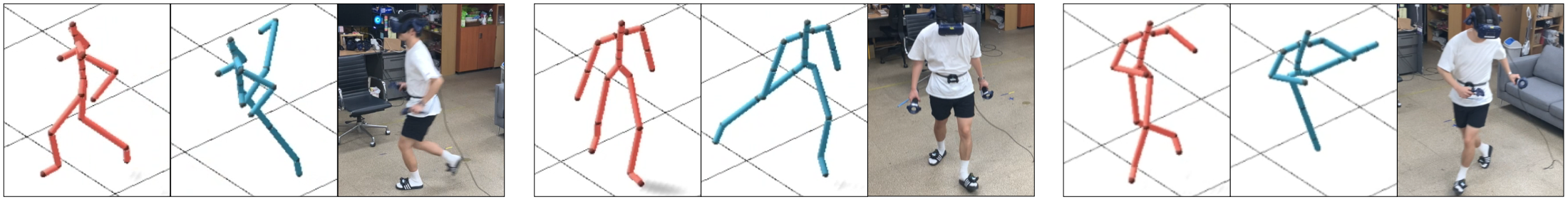}
  \vspace*{-3mm}
  \caption{\label{fig:Robustness}
           Comparison of robustness to feet tracking loss against 6-point tracking (Red: Ours, Green: 6-point tracking).}
  \vspace*{13mm}

\end{figure*}

Figure \ref{fig:DiffPeople} shows the snapshots of participants capturing various actions. Our network successfully reconstructed lower-body pose corresponding to the input upper-body motion and maintained temporal continuity between output poses. Interestingly, our model successfully learned gait patterns without using any phase or contact labels as input, matching leg movement and the feet contact states with the upper-body motion of the participants. Furthermore, our network successfully created the lower-body pose from unobserved inputs of diverse motion categories, different motion styles, and body shapes.

\paragraph*{Comparison against 6-Point Tracking.}

Figure \ref{fig:Realtime} compares avatar animation obtained by our method and by a 6-point tracking (4 trackers as ours and 2 additional trackers at feet) method. The lower body motion of the 6-point tracking method is created by inverse kinematics. One can see that the resulting animation of our method is similar to that from the 6-point tracking.

Due to the infrared occlusion by existing objects, the feet tracking often failed and caused the 6-point tracking avatar to make bizarre poses (Figure \ref{fig:Robustness}). In contrast, our method does not suffer from such tracking loss and outputs lower-body motion robustly. 

\section{Limitation and Future Work}

Our method has several limitations, which should be overcome in order to enable users to freely capture the unbounded range of high-quality motions with sparse tracking data.

First, we fixed the Y-axis of the reference joint to the world up-vector to clearly capture the ground-plane translation and rotation. As a result, the reference joint can only rotate about Y-axis but not about the frontal or lateral axes, which are major axes for some rotational actions like rolling and windmill. To deal with these rotations, an additional step of predicting frontal and lateral rotations of the pelvis by using all four tracking signals would be necessary.

Second, while our method shows competitive results for contact prediction in the majority of cases, prediction accuracy drops when the user makes sharp turns or complex leg-crossing motions, which may cause inappropriate fixation of feet or skating/floating behavior. It remains a future goal to develop more robust contact predictors and corresponding leg animation methods.

Third, the output lower-body animation sometimes contains jittering and sliding artifacts. In most cases, the artifacts are due to the unintended movement of the pelvis tracker, which is attached in front of the abdomen and thus moves differently from the actual pelvis joint. An additional source of artifact is the ambiguity occurring when a stationary user starts moving; it is difficult to determine whether the user begins locomotion or moving in-place only from past observations. For this reason, artifacts are observed during dynamic in-place motions such as making boxing or walking gestures. Utilizing short future information \cite{huang2018deepinertialposer} and exploring the autoregressive approach \cite{martinez2017human} are promising future directions to reduce this ambiguity.

Lastly, a very challenging goal with high impact is to capture the full-body motion by using only a minimal 3 trackers on the head and hands, without using the pelvis tracker, which is cumbersome to wear on top of clothing. However, considering the great amount of ambiguity in human poses given only three tracker signals, we conjecture that it would be extremely difficult to develop a 3-point tracking system that can distinguish a wide range of human poses. A feasible direction might be to combine with additional sensory modals, e.g., attaching a vision sensor on HMD, that reinforce the full-body motion capture.
\section{Conclusion}
We presented a novel method to reconstruct lower-body motion from sparse tracking data of upper-body joints, using a GRU-based network architecture. Our method does not require to use feet trackers, which are error-prone due to IR occlusion and foot-ground impact. Our GRU-based structure successfully learns the temporal characteristics of human motion. The velocity-based prediction scheme is robust against variations in tracker attachment and users' body shape. Overall, our system generates lower-body motion that is visually competitive to the sequence obtained from the baseline 6-point tracking system with additional feet trackers.

\section*{Acknowledgement}
This work was supported in part by KEIT, Korea (20011076) and Korea Creative Content Agency, Korea (R2020040211).
\vspace{\baselineskip}
\section*{Appendix}

\subsection*{1. Velocity Computation}

Let $(q^{ref}_i, p^{ref}_i)$ and $(q^{joint}_i, p^{joint}_i)$ denote the world orientation and position of the reference joint and descendent joints at the $i^{th}$ frame, respectively. Then the equations to compute velocity $\textbf{d}_{i} = [d_{i}^{ref}, [d_{i}^{joint, k}]_{k=1}^{3}]$  are given below:
\[
\begin{array}{ll}
d_{i}^{ref} & = [v_{i}^{ref}, w_{i}^{ref}] \\
v_{i}^{ref} & = \inv{(q_{i}^{ref})} * (p_{i}^{ref} -p_{i-1}^{ref}) \\
w_{i}^{ref} & = \inv{(q_{i}^{ref})} * \inv{(q_{i-1}^{ref})} * q_{i}^{ref} \\\\
d_{i}^{joint} & = [v_{i}^{joint}, w_{i}^{joint}] \\
v_{i}^{joint} & = ({p'_{i}}^{joint} - {p'_{i-1}}^{joint}), \\
p' & = \inv{(q^{ref})} * (p^{joint} - p^{ref})  \\
{w_{i}}^{joint} & = \inv{({q'_{i-1}}^{joint})} * {q'_{i}}^{joint} \\
q' & = \inv{(q^{ref})} * q^{joint}.
\end{array}
\]

\subsection*{2. Dataset Formation}

The goal of this research is to reconstruct general human motions from sparse tracking signals. We built a dataset combining parts of three different motion capture datasets. PFNN dataset \cite{holden2017phasefunctioned} is selected as the main dataset for containing various locomotion styles and transitions. We only used general locomotion animations of the PFNN dataset and excluded animations of climbing up and jumping. Selected animations consist of different motion categories, including stand, walk, jog, and run; the categories are randomly placed in an animation sequence and various transitions, such as normal turn, side steps, and back steps, exist between motion categories.

To cover in-place motions such as sit, stand up, and upper-body gestures, we selected animations in the CMU motion capture dataset  \cite{cmu2013} and MHAD dataset \cite{ofli2013berkeley}, and synthesized sequential animations for the combined dataset by connecting, clipping, and looping the selected animations.

All raw motion capture animations
were down-sampled to 45 fps and the root position and joint offsets were edited (1. scaling parameter: $0.0594$, 2. root height shift: $-0.05$m) to match the root height of $1$m in T-pose. Table \ref{tab:dataset} shows the ratio and the total number of frames of different action categories. Tables \ref{tab:training composition} and \ref{tab:test composition} show the sources and names of animations consisting of the training and test sets, respectively.

\begin{table}[ht]
\centering
\caption{Ratio of action categories in the dataset.}
\begin{tabular}{ |l|r|r|r| } 
\hline
Category & Frames & Duration (min) & Ratio \\
\hline
\hline
Locomotion & $141,340$ & $52.35$ & $84.0\%$ \\ 
Sit/Stand & $7,548$ & $2.80$ & $4.5\%$ \\ 
Upper-body motion & $19,402$ & $7.19$ & $11.5\%$ \\ 
\hline
\textbf{Total} & $\textbf{168,290}$ & $\textbf{62.34}$ & $\textbf{100}\%$ \\ 
\hline
\end{tabular}
\label{tab:dataset}
\end{table}%

\begin{table}[ht]
\footnotesize
\centering
\caption{Training set composition.}
\begin{tabular}{ |l|l|l| } 
\hline
Category & Source & Name \\
\hline
\hline
\multirow{11}{*}{Locomotion} & \multirow{11}{*}{PFNN} &NewCaptures01\_000 \\
&&NewCaptures02\_000 \\
&&LocomotionFlat01\_000 \\
&&LocomotionFlat02\_000 \\
&&LocomotionFlat03\_000 \\
&&LocomotionFlat04\_000 \\
&&LocomotionFlat05\_000 \\
&&LocomotionFlat06\_000 \\
&&LocomotionFlat07\_000 \\
&&LocomotionFlat08\_000 \\
&&LocomotionFlat10\_000 \\
\hline
\multirow{5}{*}{Sit/Stand} & \multirow{5}{*}{MHAD} & skl\_s01\_a10\&a11\_r01 \\
&& skl\_s01\_a10\&a11\_r02 \\
&& skl\_s01\_a10\&a11\_r03 \\
&& skl\_s01\_a10\&a11\_r04 \\
&& skl\_s01\_a10\&a11\_r05 \\
\hline
\multirow{2}{*}{Upper-body motion} & \multirow{2}{*}{CMU} & 15\_05 \\
&& 111\_22 \\
\hline
\end{tabular}
\label{tab:training composition}
\end{table}%

\begin{table}[ht]
\centering
\footnotesize
\caption{Test set composition.}
\begin{tabular}{ |l|l|l| } 
\hline
Category & Source & Name \\
\hline
\hline
\multirow{4}{*}{Locomotion} & \multirow{4}{*}{PFNN} 
&LocomotionFlat02\_001 \\
&&LocomotionFlat06\_001 \\
&&LocomotionFlat08\_001 \\
&&LocomotionFlat11\_000 \\
\hline
\multirow{2}{*}{Sit/Stand} & \multirow{2}{*}{MHAD} & skl\_s02\_a10\&a11\_r01 \\
&& skl\_s02\_a10\&a11\_r02 \\
\hline
\multirow{2}{*}{Upper-body  motion} & \multirow{2}{*}{CMU} & 32\_11 \\
&& 29\_18 \\
\hline
\end{tabular}
\label{tab:test composition}
\end{table}%

\textit{Tracking Signal Augmentation.}
To augment imaginary tracking signals from motion data, we first computed the world transformations for the 4 joints of the head, finger-bases, and root that correspond to Head-Mounted display (HMD), two hand-held controllers, and pelvis tracker, respectively. The VIVE Lighthouse tracking system has an expected accuracy of 2mm \cite{VIVEnoise}. To apply this drifting behavior of the trackers to the training data, we added random vectors with scale sampled from a normal distribution $\mathcal{N}(0, 0.01)$, to the position of joints. For rotational noise, random rotations with a maximum angle of 1.5 degrees were applied to the joint world rotations.

\textit{Contact Labeling.}
Each frame of a motion clip has binary contact labels for two toe-bases. We simply labeled that a toe-base is in contact if the height of the toe-base joint is below 1cm, which gave us a reasonable result for the dataset.

\subsection*{3. Details on Network Training}
The number of training epoch is $1,500$. We use Adam optimizer with an initial learning rate of $10^{-3}$ with a decaying rate of 0.999 every epoch. We use a batch size of 256; to form a batch, 256 animations are randomly selected from the set with the probability of $p_{anim}=\frac{f_{anim}}{\sum{f}}$, where $f$ is the number of frames of the animation. Finally, a 45-frame chunk is randomly picked from each selected animation to form a batch of dimension $R^{256 \times 45 \times (4 \times 9 + 1)}$.
\vspace{\baselineskip}

\printbibliography

\end{document}